\documentclass{eptcs}
\providecommand{\event}{REFINE 2011}
\usepackage{amsmath}
\usepackage{amsthm}
\usepackage{amssymb}
\usepackage{breakurl}
\usepackage{graphicx}
\usepackage{subfig}
\usepackage{stmaryrd}

\newtheorem{definition}{Definition}
\newtheorem{theorem}{Theorem}
\newtheorem{corollary}{Corollary}
\newtheorem{proposition}{Proposition}
\newtheorem{lemma}{Lemma}
\newcommand{\PAR}{\ensuremath{\mathop{\texttt{\textup{|}}}}}

\newcommand{\Hole}[1]{{-}_{#1}}
\newcommand{\defeq}{\ensuremath{\stackrel{\textup{\tiny def}}{=}}}
\newcommand{\saferef}{\ensuremath{\stackrel{\textup{\tiny safe}}{\sqsubseteq}}}
\newcommand{\vsaferef}{\ensuremath{\stackrel{\textup{\tiny safe}}{\sqsubseteq}}}

\newcommand{\vliveref}{\ensuremath{\stackrel{\textup{\tiny live}}{\sqsubseteq}}}
\newcommand{\refines}{\ensuremath{\sqsubseteq}}
\newcommand{\vrefines}{\ensuremath{\sqsubseteq}}
\newcommand{\trace}[1]{\langle #1 \rangle}
\newcommand{\tracesof}[1]{\ensuremath{\mathit{Tr}(#1)}}

\newcommand{\GrammarSep}{\,\ensuremath{\mid}\,}

\newcommand{\nest}{\ensuremath{\mbox{\LARGE .}}}
\newcommand{\tensor}{\otimes}

\renewcommand{\c}[1]{\ensuremath{\mathsf{#1}}}
\newcommand{\cZ}{\c Z}
\newcommand{\cU}{\c U}
\newcommand{\cN}{\c N}
\newcommand{\cS}{\c S}
\newcommand{\cF}{\c F}
\newcommand{\csend}{\c {send}}
\newcommand{\crecv}{\c {recv}}
\newcommand{\cnil}{\c {nil}}

\newcommand{\aC}{\mathbf{C}}
\newcommand{\aA}{\mathbf{A}}
\newcommand{\aD}{\mathbf{D}}

\theoremstyle{remark}

\title{Bigraphical Refinement\thanks{This work funded in part by the Danish Research Agency (grant no.: 2106-080046) and the IT University of
Copenhagen (the Jingling Genies project). The first author would like to thank Prof. Steve Reeves and Dr. David Streader for hosting him as a visiting researcher at the University of Waikato during the early stages of this work, and for helpful discussions during this time.
}}
\author{Gian Perrone \qquad\qquad S\o ren Debois \qquad\qquad Thomas Hildebrandt\\
\institute{Programming, Logic and Semantics Group\\
IT University of Copenhagen\\Copenhagen, Denmark}
\email{\{gdpe,debois,hilde\}@itu.dk}
}
\def\titlerunning{Bigraphical Refinement}
\def\authorrunning{G. Perrone, S. Debois \& T. Hildebrandt}

\begin{document}
\maketitle 

\begin{abstract}
We propose a mechanism for the vertical refinement of bigraphical reactive systems, based upon a mechanism for limiting observations and utilising the underlying categorical structure of bigraphs.  We present a motivating example to demonstrate that the proposed notion of refinement is sensible with respect to the theory of bigraphical reactive systems; and we propose a sufficient condition for guaranteeing the existence of a safety-preserving vertical refinement.  We postulate the existence of a complimentary notion of horizontal refinement for bigraphical agents, and finally we discuss the connection of this work to the general refinement of Reeves and Streader.
\end{abstract}

\section{Introduction}

Refinement is the process of gradually developing a specification towards a suitable implementation, through a series of steps in which more concrete entities are shown to be as acceptable as the more abstract entities preceding it in the chain of refinement steps, based upon what may be observed of these entities.  The utility of this method has been  demonstrated through many years of application in academic and industrial settings.  In this paper we attempt to bring these well-studied benefits to a new class of systems --- namely, bigraphical reactive systems.  We focus primarily on \emph{vertical refinement} \cite{bolusset2002formal}, where the aim is to relate models constructed with respect to different semantics.

A \emph{bigraphical reactive system} \cite{milner2009space,milner06pure} (BRS) is a model construction paradigm proposed by Milner and colleagues that aims to enable modelling of interactive systems within a cohesive theoretical framework.  While the primary long-term focus of bigraphs is on models of ubiquitous and context-aware systems \cite{birkedal2006bigraphical}, they have demonstrated value in other areas such as biological applications \cite{krivine2008stochastic,Damgaard:08:ALanguageForTheCell,Damgaard:08:GenericLangBioSystBasedBigraphs} and business processes \cite{HildeBrandtEtAl:2006:BPEL,zhang08bigraphical}. Bigraphical reactive systems also capture the syntactic and semantic structure of many formalisms associated with process modelling, providing a unifying meta-calculus within which to relate many of these well-developed theories.  Already encodings into various bigraphical reactive systems have been demonstrated for amongst others the $\lambda$-calculus \cite{Milner:07:LocalBigraphsAndConfluence}, CCS \cite{milner06pure}, the Mobile Ambients calculus \cite{jensen06mobile}, several variants of the $\pi$-calculus \cite{jensen06mobile,bundgaard2006typed,Elsborg:08:TypeSystForBigaphs}, Fusion Calculus \cite{GrohmannMiculan:2007:ReactSysOverDirectedBig} and Petri Nets \cite{leifer06transition}.

Bigraphical reactive systems consist of two graphs (hence the name \emph{bi}graph) modelling the orthogonal notions of \emph{locality} and \emph{connectivity} which together capture the static structure of a system, and a set of \emph{reaction rules} that may selectively rewrite portions of the bigraph in order to capture the dynamic behaviour of that system. We will introduce bigraphs and bigraphical reactive systems (assuming no prior knowledge) in Section \ref{sec:bigraphs}.

The usual notion of ``observation'' in a BRS is derived from the above notion of dynamic behaviour: a BRS gives rise to an LTS, the labels of which are simply the least context enabling reaction. The present effort towards refinement takes this connection between static structure and dynamic behaviour to heart, and attempts to short-circuit the LTS in favour of a more directly structural mechanism of refinement. This makes sense uniquely for bigraphs exactly because of the close correspondence between structure and dynamics.  The primary contribution of this paper is to introduce such a mechanism as a small step towards bringing the well-established benefits of refinement to models constructed within the bigraph formalism.  Additionally, we give a sufficient condition for an abstraction functor (Section \ref{sec:refinement}) to give rise to a safe refinement, and show that this notion of refinement corresponds with (and indeed, in part is an instance of) the general refinement of Reeves and Streader \cite{reeves2008general1,reeves2008general2}.

\subsection{Structure of the paper}

The remainder of this paper is structured as follows:  We review bigraphs (assuming no prior knowledge) in Section \ref{sec:bigraphs}. In Section \ref{sec:example} we introduce a running example that will be used to illustrate all of the concepts presented. In Section \ref{sec:refinement} we present our definition of vertical refinement for bigraphical reactive systems and show that the proposed refinement preserves safety properties with respect to the abstraction functor upon which it is parametrised. Additionally, we present a sufficient condition for an abstraction functor to give rise to a safe refinement.  Finally, in Section \ref{sec:discussion} we discuss a  candidate horizontal refinement mechanism for bigraphical agents, derived from the general refinement of Reeves and Streader \cite{reeves2008general1,reeves2008general2}, and discuss the connection of this work to general refinement.

\section{Bigraphical Reactive Systems}
\label{sec:bigraphs}

Bigraphical reactive systems is a graphical formalism emphasising the orthogonal notions of \emph{locality} and \emph{connectivity}.  A BRS is a category of bigraphs and a set of reaction rules that may be applied to rewrite these bigraphs.  We provide here a short, informal introduction to the anatomy of a BRS without assuming any prior knowledge.  For a complete treatment of bigraphs and BRSs, readers are referred to \cite{milner2009space,milner06pure}.

\subsection{Static Structure}

\begin{figure}
\centering
  \subfloat[Place Graph]{\label{fig:placegraph}\includegraphics[width=0.2\textwidth]{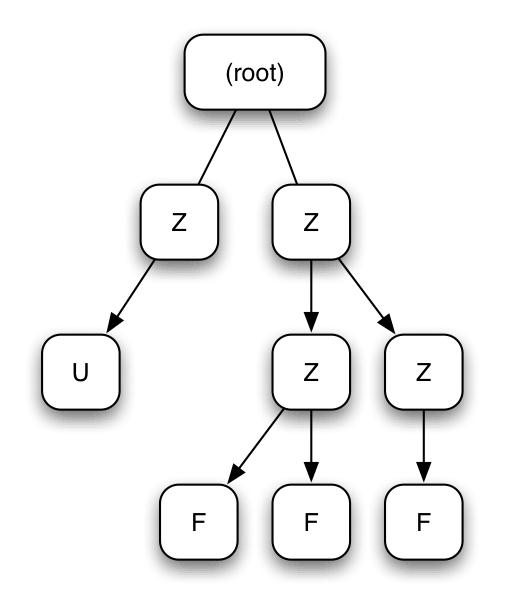}} \qquad
  \subfloat[Link Graph]{\label{fig:linkgraph}\includegraphics[width=0.3\textwidth]{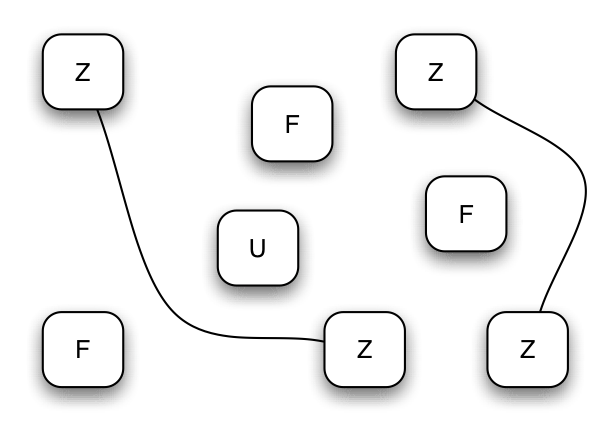}}
\caption{The constituent place (\ref{fig:placegraph}) and link (\ref{fig:linkgraph}) graphs that form a particular bigraph.}
\end{figure}

\begin{figure}
\centering
  \includegraphics[width=0.6\textwidth]{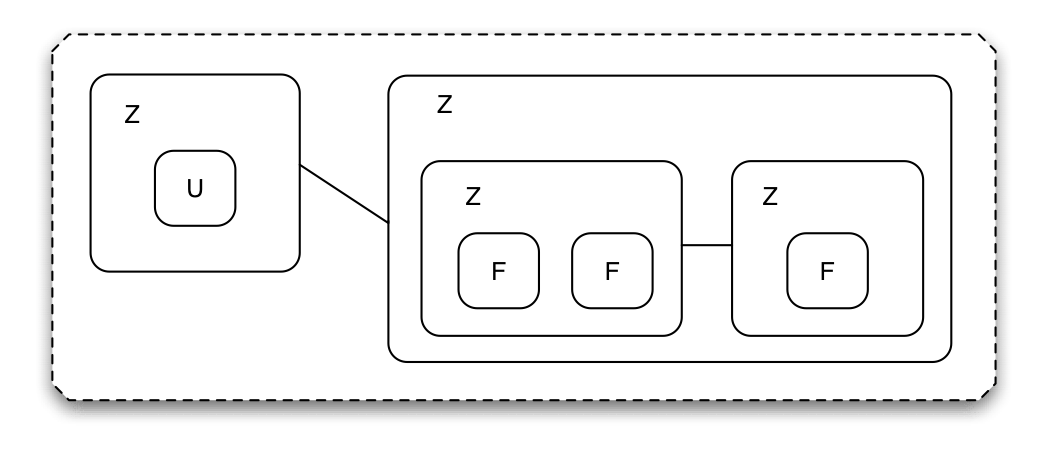}
\caption{The bigraph resulting from the combination of the place and link graphs in Fig. \ref{fig:placegraph} and Fig. \ref{fig:linkgraph}.  This bigraph is an agent of the $BRS_{notify}$ example BRS with signature $\Sigma = \{\cZ,\cU,\cF,\cN\}$ that we will introduce in Section \ref{sec:example}.}
\label{fig:model}
\end{figure}

The most basic construction within the static fragment of bigraphical reactive systems is the \emph{node}.  This follows from normal definition of a node within graph theory.  To nodes we assign \emph{controls}, which are drawn from a \emph{signature} $\Sigma$, the set of controls.   We sometimes use a convenient shorthand such that we may refer to a node as being an ``$\c X$ node'', by which we really mean a node that has been assigned the control $\c X$.  Nodes may be nested to arbitrary depth to form a tree that is known as the \emph{place graph} (Fig. \ref{fig:placegraph}).  We represent this nesting by containment, as shown in Fig. \ref{fig:model}.  We distinguish between controls of two kinds: \emph{active} and \emph{passive} ones; we shall see later how active controls admit dynamic behaviour beneath them whereas passive controls do not. Every tree of nodes is contained by a \emph{region} (the dotted border in Fig. \ref{fig:model}).  Bigraphs permit multiple regions (a place forest).

 To controls (and therefore nodes) we assign a fixed \emph{arity}, which defines the number of \emph{ports} that a given node possesses.  A port is a connection point on a node; it must always be connected to other such connection points by  the \emph{link graph}. The link graph (Fig. \ref{fig:linkgraph}) is an undirected hypergraph over the ports of the nodes of the place graph.  A single (hyper) edge may connect arbitrarily many ports on different nodes.

\begin{figure}
\centering
  \subfloat[$A : \langle 2, \{x,y\} \rangle \to \langle 1, \emptyset \rangle$]{\label{fig:composition1}\includegraphics[width=0.25\textwidth]{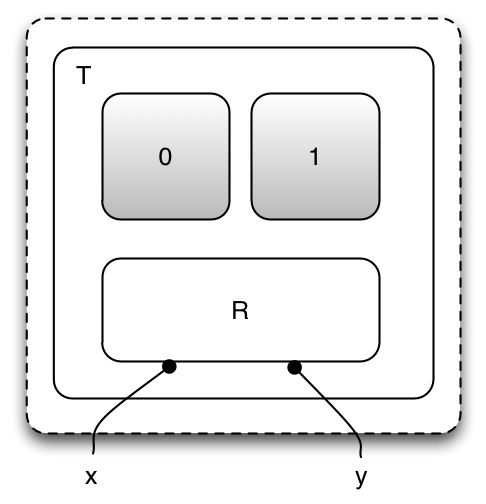}} \qquad 
  \subfloat[$B : \langle 0, \emptyset \rangle \to \langle 2, \{x,y\} \rangle$]{\label{fig:composition2}\includegraphics[width=0.33\textwidth]{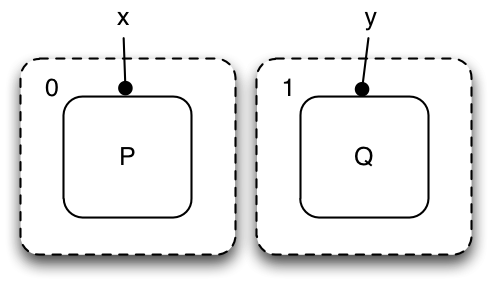}} \qquad
  \subfloat[$A \circ B : \langle 0, \emptyset \rangle \to \langle 1, \emptyset \rangle$]{\label{fig:composition3}\includegraphics[width=0.26\textwidth]{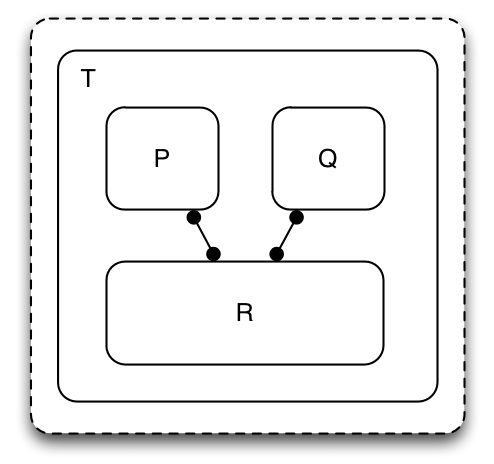}}
\caption{The composition of two bigraphs $A$ and $B$ with their respective interfaces}
\label{fig:composition}
\end{figure}

Within the place graph, in addition to regions and nodes, there may also exist \emph{holes} (known as \emph{sites} in some bigraphs literature), which are expressed visually as shaded grey nodes (as in Fig. \ref{fig:composition1}).  A hole is a location into which a region of another bigraph may be inserted by composition. It may be helpful to think of bigraphs with holes as ``contexts'' and those without as ``processes'' or ``terms''. 

Present also within Fig. \ref{fig:composition} are \emph{names} that represent (named) points at which edges of the link graph may be fused to form a single (hyper) edge. In the intuition of contexts and terms, names of bigraphs roughly correspond to unstructured names, as in the $\pi$-calculus. By convention, \emph{outer names} are drawn upwards, and \emph{inner names} are drawn downwards.  Outer names are analogous in the link graph to regions in the place graph, while inner names are analogous to holes.  Through composition of link graphs, sets of inner and outer names that agree are matched and joined.

\begin{definition}[Interface]
An interface is a pair $\langle j, X \rangle$ where $0 \leq j$, indicating the number of holes or regions, and $X$ is a set of (inner or outer) names.
\end{definition}

\begin{definition}[Bigraph]
\label{def:arrow}
A bigraph is a 5-tuple:
$$(V, E, ctrl, prnt, link) : \langle k, X \rangle \rightarrow \langle m, Y \rangle$$
Here $V$ is the set of nodes, $E$ is the set of hyperedges, $ctrl$ is the \emph{control map} that assigns controls (and therefore arities) to nodes, $prnt$ is the \emph{parent map} that defines the tree structure in the place graph and $link$ is a link map that defines the link structure.  The inner interface $\langle k, X \rangle$ indicates that the bigraph has $k$ holes, and a set of inner names $X$.  The outer interface $\langle m, Y \rangle$ indicates that the bigraph has $m$ regions and a set of outer names $Y$.
\end{definition}

\begin{definition}[Composition]
Bigraphs are composed separately in the place and the link graphs. The interfaces of the bigraphs must be compatible in order for composition to be defined, i.e., the sets of names and the number of regions/holes must be the same.  Fig. \ref{fig:composition} illustrates the composition $A \circ B$ of bigraphs $A$ and $B$.  In the place graph, we
insert contents of the left-most region of $B$ into hole 0 of $A$, and the contents of the right-most region of $B$ into hole 1
of $A$. Regions are numbered left-to-right: we insert the contents of region 0 into hole 0 etc.  In the link graph, links are spliced together where there is name agreement between the inner and outer names of the bigraphs being composed.  We may refer to $A$ in this case as being a \emph{context} into which $B$ is inserted.
\end{definition}

\begin{definition}[Tensor Product]
There exists an additional way in which to combine bigraphs, namely the \emph{tensor product} $A \otimes B$, where $A$ and $B$ are bigraphs.  Where $A$ and $B$ do not share any inner or outer names, this just involves juxtaposing their place graphs, taking the union of their names, and increasing the indices of holes in $B$ to make them unique with respect to $A$.  This definition obscures some technical details.  It is recommended that readers interested in following the proofs in Section \ref{sec:proofs} refer to \cite{milner2009space} for a precise definition.
\end{definition}

\subsection{Notation}
We introduce a rudimentary term language for representing bigraphs that should be familiar to most readers accustomed to the notation for process algebras. The present language is not complete, i.e., it cannot express every bigraph, but it can express the ones we will use in examples. It is a subset of a complete such language \cite{milner05axioms}. We will use this term language in conjunction with the graphical representation used in Fig. \ref{fig:model}.  

\begin{definition}[Bigraph Term Language]
\label{def:termlanguage}
\begin{align*}
p & ::= \kappa(n_1,\ldots,n_{ar(\kappa)}).p \GrammarSep p \PAR p \GrammarSep \Hole{i} \GrammarSep \cnil
\end{align*}
Where $\kappa \in \Sigma$.
\end{definition}

\begin{figure}
\centering
  \subfloat[$\kappa(n_1,\ldots,n_{ar(\kappa)}).p$]{\label{fig:prefix}\includegraphics[width=0.22\textwidth]{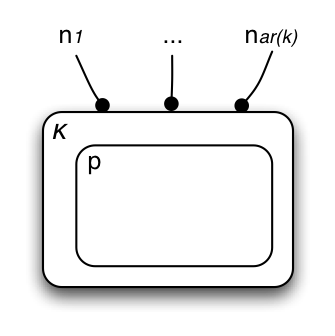}} \qquad
  \subfloat[$a.\Hole{0}$]{\label{fig:hole}\includegraphics[width=0.22\textwidth]{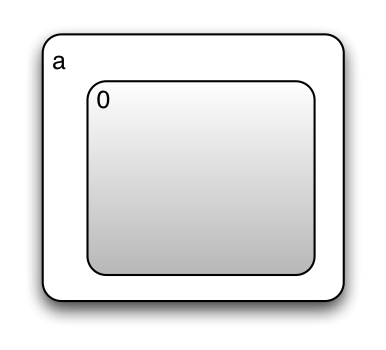}} 
  \subfloat[$a.\cnil \PAR b.\cnil$]{\label{fig:par}\includegraphics[width=0.25\textwidth]{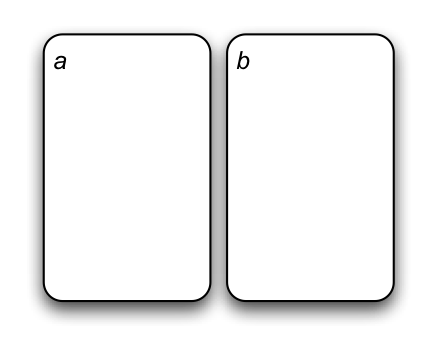}}  
\caption{Example bigraph terms with their associated graphical representation}
\end{figure}

The term language requires some explanation --- $\kappa(n1_,\ldots,n_{ar(\kappa)}).p$ is \emph{prefixing} (Fig. \ref{fig:prefix}), indicating a node assigned the control $\kappa$.  The arity of $\kappa$ is given by $ar(\kappa)$.  The sequence $n_1,\ldots,n_{ar(\kappa)}$ are the ports of the node.  Finally, the suffix $p$ is the term that is nested inside this node.  $p \PAR p$ is \emph{juxtaposition} of terms (Fig. \ref{fig:par}), placing them as siblings within the place graph.  $\Hole{i}$ is a hole (Fig. \ref{fig:hole}), indexed by some integer $0 \leq i$.  Finally, $\cnil$ is the nil terminator which is simply the empty graph in the graph representation.

\subsection{Dynamics}

Having introduced the basic structure of \emph{bigraphs}, the static portion of a BRS, we now introduce the $reactive$ portion of a BRS that imbues a system with dynamic behaviour.  This relies on \emph{reaction rules} that define rewriting that may be applied to a bigraph.  A reaction rule $(R,R',\eta)$ consists of a \emph{redex} $R$, a \emph{reactum} $R'$ and an \emph{instantiation map} $\eta$, where the redex is a bigraph to be matched and the reactum is the bigraph with which the matched portion of the bigraph should be replaced.  The instantiation map indicates how parameters matched by holes in the redex should manifest in the reactum after matching.  Where the instantiation map is unambiguous (e.g., it is the identity map), we may just write $R \to R'$.

\begin{definition}[Reaction]
Matching of a particular reaction rule $(R, R', \eta)$ against a particular bigraph $G$ and rewriting it into some other bigraph $G'$ proceeds by decomposition of the bigraph into a \emph{context} $C$, a \emph{match} $R$ (the redex), and a set of \emph{parameters} $d$ (for portions of the bigraph that are matched by holes in the redex).  This decomposition is then reassembled with the reactum $R'$ replacing the matched portion of $G$, with select parts of $d$ substituted into the holes of $R'$, forming the resulting bigraph $G'$.
$$G = C \circ R \nest d \rightarrow C \circ R'\nest \eta(d) = G'$$
We require further that the context $C$ be \emph{active}, that is, that every control above holes of $C$ are active (see CCS example below). 
\end{definition}
We have suppressed details of the handling of names here by using the notation ``$R\nest d$"; we have also suppressed details in the phrase ``with select parts of $d$" and not explained the use of the map $\eta$. We refer the reader to \cite{milner2009space} or \cite{milner06pure} for details. The present paper can be read without understanding these details, as reaction in our examples always take the form of the following special case:
\[
 	a = C \circ R\circ d \to C\circ R'\circ d\;.
\]

\begin{definition}[Bigraphical Reactive System]
We use the notation $BG(\Sigma,\cal R)$ to denote a bigraphical reactive system with a \emph{signature} $\Sigma$ (the set of constituent controls), and a set of reaction rules $\cal R$.  More formally, $BG(\Sigma,\cal R)$ is an \emph{spm category} \cite{milner2009space} in which the objects are interfaces and the arrows are bigraphs (which we refer to as \emph{agents} of $BG(\Sigma,\cal R)$), equipped with a set of reaction rules $\cal R$.
\end{definition}

As an example, we introduce a very simple calculus in the style of the Calculus of Communicating Systems (CCS) \cite{milner1980calculus}, where we first give an encoding of the terms as bigraphs, and then define a reaction rule that imbues these terms with dynamic behaviour.  Interested readers are referred to \cite{milner2009space} for a real encoding of CCS.

\begin{figure}
\centering
\includegraphics[width=0.4\textwidth]{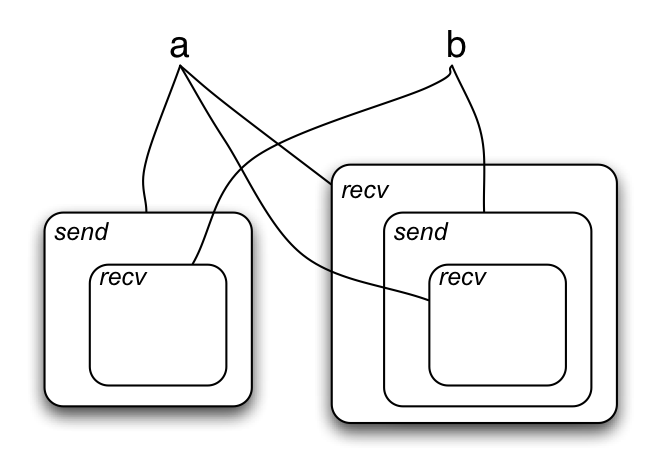}
\caption{The process $\csend(a).\crecv(b).\cnil \PAR \crecv(a).\csend(b).\crecv(a).\cnil$}
\label{fig:ccsterm}
\end{figure}

\begin{figure}
\centering
\includegraphics[width=0.6\textwidth]{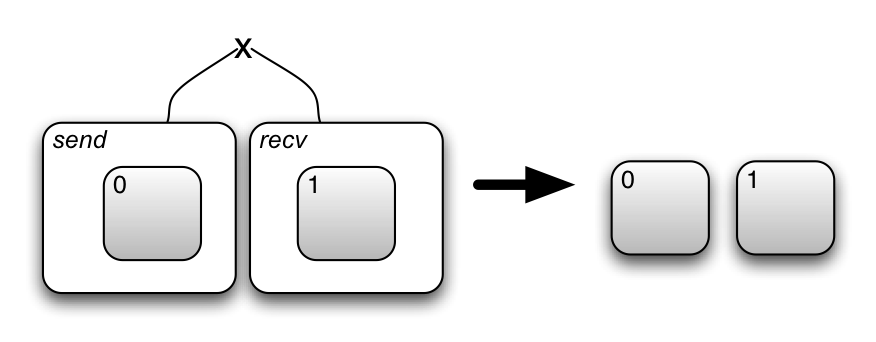}
\caption{The $R_{CCS}$ reaction rule}
\label{fig:ccsrule}
\end{figure}

Our calculus defines sequencing ($t.P$), parallel composition ($t \PAR t$), and sending and receiving on a named channel (``$x!$'' and ``$y?$'', respectively, where $x$ and $y$ are channel names).  The encoding of these constructs into the bigraphical term language in Definition \ref{def:termlanguage} is straightforward --- these primitives are already defined in terms of the bigraphical term language, except for ``send" and ``receive" which we straightforwardly encode as nodes with controls $\csend$ and $\crecv$, each with arity $1$.  Fig. \ref{fig:ccsterm} gives a graphical representation of the process $\csend(a).\crecv(b).\cnil \PAR \crecv(a).\csend(b).\crecv(a).\cnil$.  According to our encoding, sequencing is represented by prefixing, parallel composition by juxtaposition, actions (such as $\csend$ and $\crecv$) by \emph{passive} controls, and channels by outer names.  This is by no means the only encoding possible, but this technique is one of the most straightforward.

Having developed the encoding of our calculus within bigraphs, we can give a reaction rule $R_{CCS}$ that will (through repeated rewriting) reduce the term as far as possible based upon agreement between parallel processes as to which action should be taken next:

$$R_{CCS} \defeq \crecv(x).\Hole{0} \PAR \csend(x).\Hole{1} \rightarrow \Hole{0} \PAR \Hole{1}$$

This rule is presented graphically in Fig. \ref{fig:ccsrule}.  It essentially ``peels off'' the outer layers of the terms where a $\csend$ and a $\crecv$ action are linked to the same channel name, rewriting the entire bigraph to the juxtaposition of whatever was nested inside those $\csend$ and $\crecv$ controls (i.e. the parts of the bigraph matched by the holes in the redex). As an example, the CCS reaction $a!.b? \PAR a?.c! \to b? \PAR c!$ becomes the bigraphical reaction
\[
  \csend(a).\crecv(b).\cnil  \PAR \crecv(a).\csend(c).\cnil \to
  \crecv(b).\cnil \PAR \csend(c).\cnil
\]

%

\section{Example}
\label{sec:example}

Aside from their role as a meta-calculus for the study of process modelling formalisms, bigraphical reactive systems are intended to provide a basis upon which to construct models of the kinds of context-aware and ubiquitous systems that are becoming increasingly popular.  Consequently, we introduce an example based on modelling a context-aware social network notification system, such that a user is notified whenever a friend is in the same physical location.  

We will give this example without using the link-graph part of bigraphs to keep it simple. We emphasise that the example generalises to a more interesting one in which connectivity counts --- where notification is dependent not only on physical co-location but also on whether or not users and friends are virtually connected through their laptops and phones.  

We will subsequently extend this to a system in which not all friends, but rather only particular designated ``special friends'', trigger notifications, and show that (and in what sense) the latter system is a refinement of the former.

The example system captures the dynamics of some physical environment (consisting of discrete zones within which we can detect the presence of a user by some mechanism that is outside the scope of this model) in which a user's friends move from zone to zone.  When one of the user's friends is present in the same zone as the user, a notification is given, modelled by  adding a ``notification'' node to the zone.

\subsection{The abstract system: $BRS_{notify}$}

We first define controls $\cZ$ (Zone), $\cU$ (User), $\cF$ (Friend), $\cN$ (Notification) and $\cS$ (Special friend marker). Every control has arity 0 and every control is active; altogether we have a signature
\[
\Sigma_N = {\cZ,\cU,\cF,\cN}
\] The bigraphs of our systems are thus arbitrary trees over these controls. We shall of course be interested only in those where $\cZ$ are inner nodes and the remaining controls are leaves. 

\begin{figure}[tbh]
\centering
  \subfloat[$M_1$]{\label{fig:move1}\includegraphics[width=0.55\textwidth]{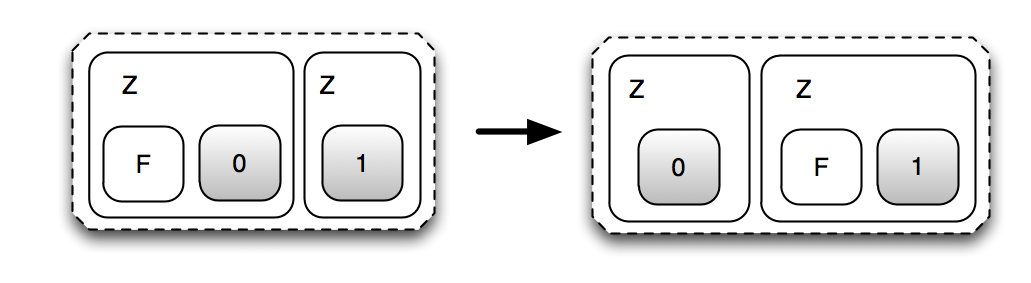}}  \\
  \subfloat[$M_2$]{\label{fig:move2}\includegraphics[width=0.55\textwidth]{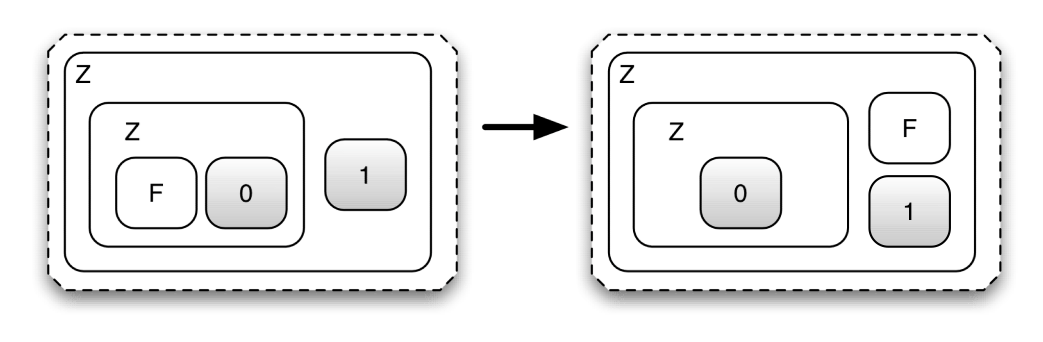}} \\
  \subfloat[$M_3$]{\label{fig:move3}\includegraphics[width=0.55\textwidth]{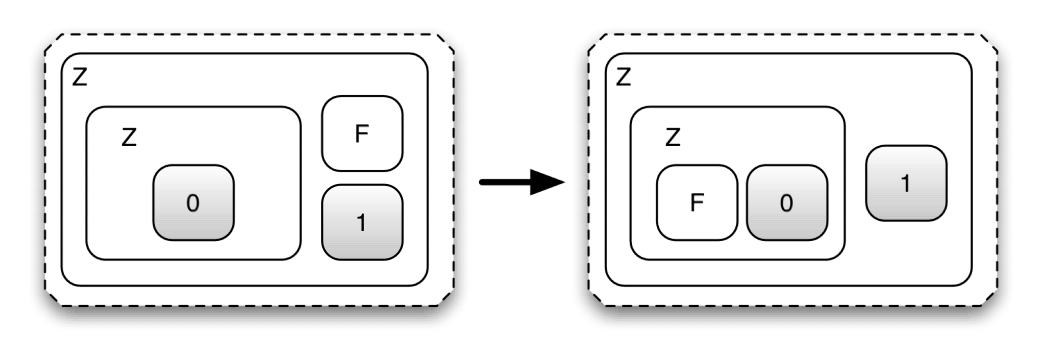}}  
\caption{Reaction rules $M_1$, $M_2$ and $M_3$ that allow $friend$ nodes to move between $zones$.}
\label{fig:movement}
\end{figure}

With these particular bigraphs in mind, we give reaction rules reconfiguring a bigraph by allowing nodes with control $\cF$ --- friends --- to move between nested zones as follows. These rules are illustrated graphically in Fig. \ref{fig:movement}.
\label{def:movement}
\begin{align*}
M_1 & = \cZ.(\cF \PAR \Hole{0}) \PAR \cZ.\Hole{1}  \qquad \rightarrow\quad 
\cZ.\Hole{0} \PAR \cZ.(\cF \PAR \Hole{1})
			 \\
M_2 & = \cZ.(\cZ.(\cF \PAR \Hole{0}) \PAR \Hole{1}) \; \quad \rightarrow\quad  
			 \cZ.(\cZ.\Hole{0} \PAR \cF \PAR \Hole{1})  \\
M_3 & = \cZ.(\cZ.\Hole{0} \PAR \cF \PAR \Hole{1}) \qquad \rightarrow\quad 
			 \cZ.(\cZ.(\cF \PAR \Hole{0}) \PAR \Hole{1}) 
\end{align*}
Reaction rules are here given on the form ``$R\to R'$'' rather than the more precise $(R,R',\eta)$; recall from the above introduction to bigraphs that we use the former form whenever $\eta$ is inconsequential (in this case, it is the identity map).

\begin{figure}
\centering
  \includegraphics[width=0.7\textwidth]{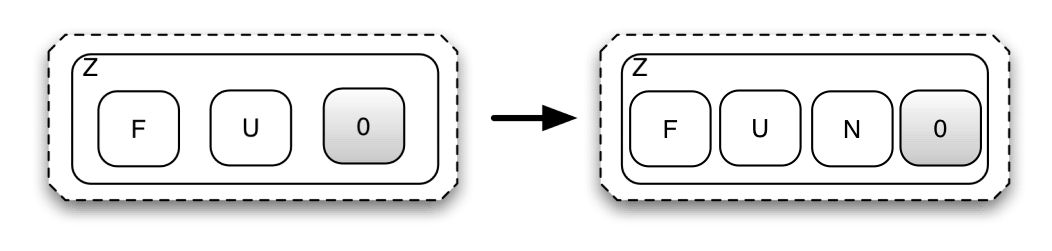}
\caption{Reaction rule $R_1$}
\label{fig:notify1}
\end{figure}

We extend the movement rules $M$ with an additional rule $R_1$ for notifications to be issued when a $\cU$ (user) and $\cF$ (friend) node exist within the same zone.  This reaction rule is illustrated in Fig.~\ref{fig:notify1}.  
\begin{align*}
\Sigma_N & = \Sigma_M \cup \{\cU, \cN\} \\
R_1 & = 
         \cZ.(\cU \PAR \cF \PAR \Hole{0}) \qquad \rightarrow \qquad
			 \cZ.(\cU \PAR \cF \PAR \cN \PAR \Hole{0})
\end{align*}

Let $BRS_{notify}$ be the bigraphical reactive system formed by the addition of the reaction rule $R_1$ to the set of movement rules $M$:

\begin{align*}
BRS_{notify} & = BG(\Sigma_N, M \cup \{R_1\})
\end{align*}

\subsection{The concrete system: $BRS_{selective}$}

\begin{figure}
\centering
  \includegraphics[width=0.6\textwidth]{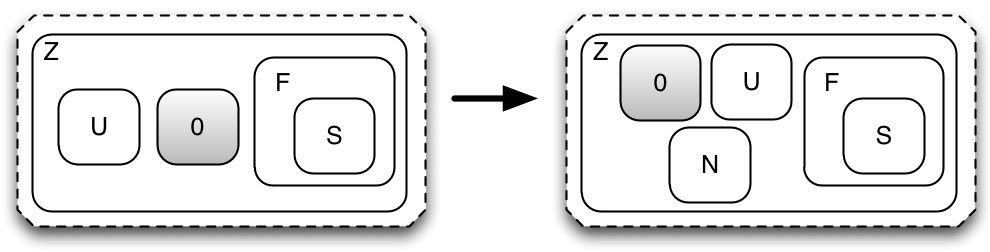}
\caption{Reaction rule $R_2$}
\label{fig:notify2}
\end{figure}

We now create a second bigraphical reactive system, this one refining (both intuitively and in a sense to be made precise) the system $BRS_{notify}$ just introduced. In this new system,
 instead of simply notifying whenever \emph{any} friend is present in the same zone as the user, we wish only to issue a notification in the presence of a particular designated friend, distinguished by the presence of an $\cS$ (special friend marker) inside the friend node in question.  Consequently, we define the set of controls $\Sigma_{\cS}$ for $BRS_{selective}$ to include (in addition to the controls of $\Sigma_N$) the $\cS$ control.  The modified reaction rule $R_2$ is presented graphically in Fig. \ref{fig:notify2}.
\begin{align*}
\Sigma_S & = \Sigma_N \cup \{\cS\} \\
R_2 & = 
         \cZ.(\cU \PAR \cF.\cS \PAR \Hole{0}) \rightarrow
			 \cZ.(\cU \PAR \cF.\cS \PAR \cN \PAR \Hole{0})\\
BRS_{selective} & = BG(\Sigma_S, M \cup \{R_2\})
\end{align*}
At an intuitive level, this BRS refines the one of the previous sub-section. In the following section, we shall define exactly in what sense this is the case.

\section{Vertical BRS Refinement}
\label{sec:refinement}

We recall the distinction here between \emph{horizontal} and \emph{vertical} refinement.  Vertical refinement is concerned with moving between differing levels of abstraction, or indeed completely independent modelling languages, whereas horizontal refinement instead aims to relate models specified at the same fundamental level of abstraction, and within the same modelling setting.  When we refer to the refinement of a BRS, we mean a vertical refinement, indeed, this is the only meaningful interpretation, as a BRS is the category consisting of (infinitely) many actual agents of the same general shape.  We will later return (briefly) to what it would mean for an \emph{agent} to be refined, that is, to a horizontal refinement between two agents of the same BRS (each of which would be bigraphs, representing --- for example --- two CCS processes).  

To summarise the distinction between horizontal and vertical refinement in the setting of BRSs: In the former case, we are talking about what we can observe of all such agents, whereas in the latter we are referring to what we can observe of the behaviour of a single agent. In the present section, we consider vertical refinement; we comment on horizontal refinement in the subsequent section.

\subsection{Safe refinements}
\label{sec:proofs}

First, what observations can you make of bigraphical agents? While the notion of a trace is familiar within refinement literature, within bigraphical reactive systems it is unclear exactly what might correspond to an \emph{action} within the usual definition of a trace.  Consequently, we formulate a trace of a BRS such that each element of the trace is a bigraphical agent (i.e., a bigraph of that BRS).  Therefore the notion of trace is not one of a system exhibiting behaviour in the form of some observable actions, rather, it is the entire state of the model as it changes over time such that every element of the trace is a bigraph, related to the next element of the trace by the application of some reaction rule. While this may seem very crude at first glance, it is important to remember that the dynamic behaviour of a bigraph is derived from reaction rules and the structure in a perhaps more direct manner than in many other calculi. As such, it makes sense to consider the abstract specification to comprise, by itself, an entire observation --- cf. the structure of agents of $BRS_{notify}$ above.

If an observation is a complete agent of the abstract specification, what then is an observation of an agent of the concrete system? We leave that to the system constructor, merely insisting that the observations one makes of concrete implementation agents must somehow be a function of their structure. Thus, observations of concrete agents are given by a structure-preserving map from concrete agents to abstract ones. In the parlance of category theory, this is called a ``functor'', a functor that we shall in this instance call an \emph{abstraction functor}. 

\begin{definition}[Trace, observation]
For a given BRS $A$, a \emph{trace} is a (possibly infinite) sequence of bigraphs (agents) $\trace{a_1,a_2,\ldots}$, such that for each $a_i$ and $a_{i+1}$ in the sequence there is a reaction 
$a_i \rightarrow a_{i+1}$.
If $s=\trace{s_1,\ldots, s_n}$ and $t=\trace{t_1,\ldots}$ are traces and $s_n\rightarrow t_1$, we may form the \emph{composite trace} $s;t=\trace{s_1,\ldots,s_n,t_1,\ldots}$. In this case we say that $t$ is an \emph{extension} of $s$. We write $\tracesof{A}$ for the set of all traces of a given BRS $A$. If $F:A\to A'$ is a functor and $\trace{a_1,a_2,\ldots}\in\tracesof{A}$ is a trace of $A$, we apply $F$ pointwise to obtain a trace $F(t)=\trace{F(a_1),F(a_2),\ldots}$. 
\end{definition}
Note that $\tracesof{x}$ is by definition prefix-closed; that is, for any trace $t\in\tracesof{x}$, every prefix $t'$ of $t$ is also in $\tracesof{x}$. 

Of course, not just any functor will do: to have a refinement, the dynamic behaviour of the concrete implementation must be allowed by the dynamic behaviour the abstract specification allows on its agents, the observations. Altogether, our notion of refinement follows from the usual trace equality, however, because a BRS tends to permit too much observation, our bigraphical notion of refinement requires as a side condition that there exist an abstraction functor $F : C \to A$ such that for any trace $\trace{c_0, c_1, \ldots}$, $F$ gives rise to a trace $\trace{F(c_0), F(c_1), \ldots}$.  We present vertical refinement as the conjunction of two constituent definitions, separating the preservation of orthogonal safety and liveness properties through refinement.

\begin{definition}[Safe Vertical Refinement]
\label{def:saferefinement}
$$A \vsaferef_F C\quad \defeq\quad F(\tracesof{C}) \subseteq \tracesof{A}$$
\end{definition}
This definition satisfies the ``reduction of nondeterminism'' role of refinement, in that it is always valid to simply pick one alternative and implement it in $C$ when presented with nondeterministic choice in $A$.

\begin{lemma}
Safe Vertical Refinement is transitive and reflexive for the identity functor.
\end{lemma}
\begin{proof}
Reflexivity is trivial. Suppose $A \vsaferef_F C$ and $C \vsaferef_G D$. Then $FG(\tracesof{D}) \subseteq F(\tracesof{C})\subseteq \tracesof{A}$.
\end{proof}

We proceed to illustrate safe refinement using the two BRSs above, then give a sufficient condition for an abstraction functor to yield a safe refinement. 

Recall our claim that $BRS_{selective}$, which issues notifications upon co-location with ``special friends" is a refinement of $BRS_{notify}$, which does so upon co-location with any friend. The latter employs an additional control $\cS$.  This indicates that our abstraction functor must (at the very least) ensure that all nodes of control $\cS$  must be hidden, renamed or removed so as to ensure that the codomain of $F$ is $BRS_{notify}$ (i.e. that $F$ can transform any agent of $BRS_{selective}$ into a valid agent of $BRS_{notify}$).

By this reasoning, we arrive at an abstraction functor ``pattern'' that is likely applicable to many other BRSs.  We call this the \emph{hiding functor}.  Its essential function is to simply hide, for a given signature $\Sigma$, all nodes that have been assigned controls from some particular set of controls $H$.  This includes joining any children of nodes that will be hidden to parents that will remain visible after the application of the hiding functor.  For our example, the hiding set $H = \{\cS\}$ (i.e. the designated ``special'' friend control). 

\begin{definition}[Hiding Functor] We define an abstraction functor $F_{\Sigma,H} : BG(\Sigma) \rightarrow BG(\Sigma \setminus H)$ for hiding, parametrised by $\Sigma$, the signature of the ``implementation'' BRS, and $H$, a set of controls to be hidden. On \emph{objects}, this functor is the identity. On \emph{arrows}, its action is 
$
F_{\Sigma,H}((V,E,prnt,ctrl,link)) \defeq (V',E,ctrl',prnt',link)$, where
\renewcommand{\labelitemi}{--}
\begin{itemize}
\item $V'  = \{v \in V | ctrl(v) \notin H\} $ 
\item $ctrl'  = ctrl \downharpoonright V' $, and
\item $
prnt'(l)  \defeq
\begin{cases}
prnt(l) & \text{where $ctrl(prnt(l)) \notin H$} \\
prnt'(prnt(l)) & \text{otherwise} 
\end{cases}$
\end{itemize}
\end{definition}

This ``hiding functor'' is an abstraction functor for our example system.  Recalling the definition of a bigraphical agent (and therefore of an arrow in the category $BRS_{notify}$ or $BRS_{selective}$) given in Definition \ref{def:arrow}, the purpose of this hiding functor is to exclude any nodes that have a control that is in the set of hidden controls $H$, exclude these controls from the control map $ctrl$, and recursively recreate the parent map $prnt$ such that any children of a node with a control in $H$ is attached to its most immediate place-graph ancestor that is not marked with a control in $H$.  We call the abstraction functor for our example notification system $A_{friend}$, which is defined as the hiding functor above, instantiated with $H = \{\cS\}$. 

While the hiding functor has the flavour of a forgetful functor --- it dispenses with structure --- it cannot reasonably be called so as it is not faithful. Many distinct configurations (e.g. special-friend controls) will map to the same bigraph.  This is a technical distinction only; we use ``hiding'' in no special sense, except as a name for abstraction functors of this general shape.

It is easy to prove that with $A_{friend}$ as abstraction functor, $BRS_{selective}$ is indeed a safe refinement of $BRS_{notify}$. However, instead of proving so directly, we shall instead provide a general theorem about abstraction functors: When they preserve reaction, and in particular, when they preserve just reaction \emph{rules}, they give rise to safe refinement. 

\begin{theorem}
Let $F:C \to A$ be an abstraction functor. If $F$ preserves reaction, that is, if $c\to c'$ implies $F(c)\to F(c')$, then $A\vsaferef_F C$.
\end{theorem}
\begin{proof}
Immediate from Definition \ref{def:saferefinement} of safe refinement.  
\end{proof}

From this theorem it becomes apparent that an abstraction functor may be any functor at all that obeys this property.

The terminology deceives, here: The guarantee that the concrete system has \emph{no more} behaviour than the abstract one is in fact upheld by the abstraction functor \emph{preserving} behaviour. 

Of course, proving that a functor preserves reaction need not at all be easy. Fortunately, we can exploit the connection between static structure and dynamic behaviour of bigraphs: a functor which preserves the reaction \emph{rules}, structurally, will also preserve (dynamic) reaction, and will thus be a safe refinement.  

\begin{theorem}[Safe Abstraction Functors]
Let $A = BG(\Sigma, \cal R)$ and $C=BG(\Sigma', \cal R')$ be BRSs. A functor $F:C\to A$ yields a safe vertical refinement $A\vsaferef_F C$ \emph{if} it satisfies the following conditions. 
\begin{enumerate}
\item It preserves and respects tensor.
\item It preserves active contexts.
\item It preserves reaction rules: For any reaction rule $(R,R',\rho)\in\cal R'$ (a)
the $F$-image $(F(R), F(R'),\rho)$ is a rule in $\cal R$; and
(b) for any parameter $d$ of that rule, $\overline\rho(F(d))=F(\overline\rho(d))$.
\end{enumerate}
\end{theorem}
\begin{proof}
Suppose $c_1,\ldots,c_n$ is a trace of $C$. It is sufficient to prove that for each $i<n$, there is a reaction $F(c_i)  \to F(c_{i+1})$. We know that $c_i \to c_{i+1}$, so there is some reaction rule $(R,R',\rho)\in \cal R'$, context $E$ of $C$, and some set of names $\cZ$  s.t.
\[
   c_i = E\circ (R\tensor 1_Z) \circ d\qquad \to \qquad E\circ (R'\tensor 1_Z)\circ\overline\rho(d) = c_i'
\]  
Where $\overline\rho(d)$ is the instantiation of parameters (see \cite{milner2009space} for details).  But then, because $(F(R),F(R'),\rho)$ is a rule of $\cal R$, we 
compute and find $	a_i = F(c_i)
		 =  F(E\circ (R\tensor 1_Z) \circ d) 
	       =  F(E)\circ (F(R)\tensor 1_{F(\cZ)}) \circ F(d) 
	       	\to F(E)\circ (F(R')\tensor 1_{F(\cZ)}) \circ \overline\rho(F(d)) 
		= F(E)\circ (F(R')\tensor 1_{F(\cZ)}) \circ F(\overline\rho(d)) 
		= F(E\circ (R'\tensor 1_Z)\circ \overline\rho(d))
		= F(c_i') 
		= a_i'
$
\end{proof}

We remark that the three conditions of this Theorem appear to be good candidates for a definition of a morphism of \emph{parametric} reactive systems, as suggested in the forthcoming \cite{debois11calculation}.

It is straightforward to verify that for our example BRSs, $BRS_{selective}$ and $BRS_{notify}$, the hiding functor does in fact satisfy the three conditions of this Theorem. Thus we have the following corollary:

\begin{corollary}
$BRS_{selective}$ is a sound refinement of $BRS_{notify}$ with respect to the abstraction functor $A_{friend}$, that is, 
\[
BRS_{notify}\vsaferef_{A_{friend}}BRS_{selective}
\]
\end{corollary}

The $\saferef$ relation captures safety properties of the system being refined (i.e. it does not permit a refined model any undesirable extra behaviour, provided that the abstraction functor does not hide any ``undesirable'' behaviour).  However, it does not guarantee that the system does anything at all (i.e. an empty trace is a safe refinement of any system).  To guarantee that some additional liveness properties are preserved by refinement, it is necessary to extend our definition.

\subsection{Live refinements}

In order to guarantee that a given concrete system actually exhibits any of the desirable behaviour of the abstract system that it refines, we must define a notion of liveness.  Whereas in a process algebraic setting it might be possible to rely on the presence of a particular output (or all possible outputs) to define ``desired'' observable behaviour, within a bigraphical setting the lack of any primitive notions of ``input'' or ``output'' (it is up to the system designer to define what these concepts mean with respect to a particular model) means that it is necessary to explicitly choose such ``desirable'' behaviours.

In the absence of an intrinsic notion of desirable behaviour, we further  parametrise our notion of liveness, already parametric in terms of the abstraction functor $F$, on the admissible traces.  This parametrisation on the notion of admissibility is akin to those used in \cite{Hildebrandt99categoricalmodels,Hennessy85thepower}.

\begin{definition}[Live Vertical Refinement]
\label{def:liveref}
Let $F:C\to A$ be an abstraction functor, let $\aC\subseteq\tracesof{C}$ be the \emph{admissible traces} for $C$, and let similarly $\aA\subseteq\tracesof{A}$, the admissible traces of $A$.
We then say that $(C,\aC)$ is a \emph{live} refinement of $(A,\aA)$ iff for every trace $s$ of $\tracesof{C}$, whenever $F(s)$ has an extension $t'$ to an admissible trace $F(s);t'\in\aA$, then there exists an extension $s'$ of $s$ to an admissible trace $s;s'\in \aA$ with $F(s')=F(t')$. In this case we write:
\[ (A,\aA)\vliveref_{F} (C,\aC)\;. \]
\end{definition}

If we wish to take the admissible traces $\aA$ of the abstract system $A$ as canonical, we can define $\aC$ as those traces whose $F$-images are admissible.

\begin{lemma}
Live Vertical Refinement is transitive.
\end{lemma}
\begin{proof}
 Suppose $ (A,\aA)\vliveref_{F} (C,\aC)$ and $(C,\aC)\vliveref_{G} (D,\aD)$, and suppose $FG(t);u'\in \aA$. Then $u'$ has a pre-image $s'$ with $G(t);s'\in\aC$; but then $s'$ has a pre-image $t'$ with $t;t'\in\aD$. 
\end{proof}

Let us provide a suitable set of admissible traces for our running example, $
BRS_{notify}$. For this BRS, the obvious notion of admissibility (think ``successful'') is when notification has occurred. So we define the set of admissible traces as simply those finite traces in which the user has been notified, that is, in which the final agent contains the notification control next to the user and his friend:
\[
\mathbf S_{notified} \quad\defeq\quad
\{ \trace{a_1,\ldots, a_n} \in\tracesof{BRS_{notify}} \mid 
   \exists C.\;\; a_n = C\circ (\cU\PAR \cF\PAR \cN) \}
\]
For $BRS_{selective}$, we transfer the notion of admissiblity:
\[
\mathbf S_{selective} \quad\defeq\quad
\{ t \in \tracesof{BRS_{notify}} \mid
   F(t)\in\mathbf S_{notified}
   \}
\]
The selective system $BRS_{selective}$ under these notions of admissibility is in fact \emph{not} a live refinement of the original one $BRS_{notify}$. One might think so: After all, one can extend a trace to admissibility simply by moving the special friend next to the user. Unfortunately, there need not be a special friend, and even if there were, the abstract system might extend to admissibility by moving a (non-special) friend next to the user. We will now show this in detail, thus proving of the following proposition:

\begin{proposition}
$(BRS_{notify},\mathbf S_{notify})\not\vliveref_{A_{friend}}(BRS_{selective},\mathbf S_{selective})$.\end{proposition}
\begin{proof}
Consider an agent $\cZ.(\cU\PAR \cF)$ of $BRS_{selective}$. Applying $A_{friend}$ we find simply $A_{friend}(\cZ.(\cU\PAR \cF))=\cZ. (\cU\PAR \cF)$, which succeeds after just one reaction 
\[
\cZ.(\cU\PAR \cF) \to 
\cZ. (\cU\PAR \cF\PAR \cN)
\]
by reaction rule $R_1$. Now, if we actually had a live refinement, we should be able to match this reaction in $BRS_{selective}$. A simple inspection of the rules however prove that this is not possible.
\end{proof}

This is, however, not a show-stopper, rather it is a welcome demonstration of the utility of such a vertical refinement mechanism.  We could remedy this situation by introducing into $BRS_{selective}$ an additional reaction rule that spontaneously adds the designated friend marker $\cS$ to any friend $\cF$. However, this seems to contradict the intuition of the model, so in this instance it is perhaps better to leave $BRS_{selective}$ unmodified and accept that there are (known) conditions under which this BRS cannot progress to a successful state.

Having defined our two separate (live and safe) refinement relations, we can complete the definition of safe and live vertical refinement:

\begin{definition}[Safe and Live Vertical Refinement]
$$(A,\aA) \vrefines_{F} (C,\aC)\quad \defeq\quad A \vsaferef_F C \;\land \; (A,\aA) \vliveref_{F} (C,\aC)$$
\end{definition}

\section{Discussion \& related work}
\label{sec:discussion}

Having introduced our notion of vertical BRS refinement and shown the conditions under which it is safe and live with respect to the chosen abstraction functor, we now 
discuss potential approaches to horizontal refinement and related work. As it happens, both topics take us to the general refinement of Reeves and Streader~\cite{reeves2008general1,reeves2008general2}. 

General horizontal refinement recognises three components to refinement: \emph{entities} $E$, i.e., the specifications and implementations being refined; \emph{contexts} $\Xi$, which are the environment within which the entities interact; and a \emph{user}, which defines the possible observations $O(-)$ that can be made of an entity within a particular context. Refinement is then the relation
\begin{align*}
A \refines_{\Xi,O} C \defeq \forall x \in \Xi. O([C]_x) \subseteq O([A]_x)\;,
\end{align*}
where $\Xi$ is the set of contexts, $O$ is a map assigning observations to entities in contexts, and $[-]_x$ inserts an entity into context $x$.  

Interestingly, our proposed notion of bigraphical \emph{vertical} refinement falls under the umbrella of general \emph{horizontal} refinement. Entities would be BRSs (like $BRS_{notify}$ and $BRS_{selective}$); contexts $\Xi$ would be just the trivial context, which leaves the entity unchanged. Finally, the observation map $O$ is in our case simply $Tr(-)$, the map that takes a BRS to the traces observable of it. We do not think this is a coincidence. It seems intuitive that horizontal refinement of an entire class of agents would correspond to vertical refinement.

What about general \emph{vertical} refinement, then?
The definition of vertical refinement within the general refinement framework \cite{reeves2008general2} relies upon a notion of \emph{layers}, representing a level of abstraction in terms of $(E_L,\Xi_L,O_L)$, where $E_L$ is a set of entities, $\Xi_L$ is a set of contexts and $O_L$ is an observation function.  Vertical refinement is then defined in terms of a Galois-connection that interprets high-level entities as low-level ones and vice versa. 

The analogy of this notion with our use of an abstraction functor $F : C \to A$ should be apparent: If we could find that functor $F$ to be one of an adjoint pair, we would be in an analogous situation. Unfortunately, it remains unclear if such an adjunction would retain the intuition behind the Galois-connection of general vertical refinement: morphisms (i.e., bigraphs) do not measure approximation; they represent the agents under investigation. In particular, the hiding functors used for the example in the present paper do not appear to be part of adjoint pairs. 

Leaving vertical refinement behind, what is then a good notion of horizontal refinement for bigraphs? Returning to general horizontal refinement, bigraphs actually do come with a notion of entity, context, and observation, namely agents (roughly, bigraphs with no holes/inner names), bigraph contexts (bigraphs with holes/inner names), and an LTS (given a BRS). We have in the present paper by-passed the LTS as the notion of observation, following the bigraphical connection by structure and dynamics to its extreme conclusion, using the structure of the abstract specification as the observations. 

For horizontal refinement, this approach appears not sensible: We would after all be relating agents of the same BRS. Important examples (like CCS-process refinement) cannot be expressed within this particular approach, which should guide the development of other horizontal refinement strategies for bigraphical agents. One obvious choice seems now to be the LTS intrinsic to BRSs. We have yet to pursue this option; we caution that while BRS LTSs have been successful in recovering semantics of various process algebras and other models of concurrency, it has been less successful in providing useful semantics for pervasive systems, one of our key interests. 

However, even leaving the question of suitable observations open, we would likely find a notion inside general horizontal refinement by taking
\begin{align*}
a \refines_O c \defeq \forall x \in \Xi. O(x \circ c) \subseteq O(x \circ a)\;,
\end{align*}
where $a$ and $c$ are agents of some BRS $B$; $\Xi$ is the set of contexts of that BRS, and $O$ is some notion of the semantics of agents of $B$, perhaps traces of the LTS of $B$, or perhaps some other notion.  Indeed, early indications are that this approach would be promising in recovering (for example) CCS process refinement, contingent upon an appropriate notion of observation.

\subsection{Related Work}

Restricting the set of controls admissible under a certain control, or requiring a control to be present is well-studied in bigraphs (e.g., \cite{birkedal2008construction,milner2009space,milner06pure,conchuir2009kind}).  However, that study has invariably focused on ensuring that the bigraphical LTS theory is retained under such additional constraints, and are thus only superficially related to the present paper.

Goldsmith \& Creese \cite{goldsmith2010refinement} explore an approach to refinement within bigraphs (and particularly within Spygraphs, a specialisation of bigraphs).  They observe the ease with which one may derive an LTS for a BRS that is labeled exclusively by the trivial context $id$ (equivalent to a $\tau$ action in a process algebraic setting).  These kinds of contextual labels are not helpful for analysis, as they capture no behaviour.  Similarly, the LTS semantics of bigraphs share the same intentionality inherent in the graphical presentation.  While Goldsmith \& Creese suggest (to good effect in a CSP setting) that it may be appropriate to perform hiding at a process-level before considering a transition into bigraphs, this would seem inappropriate for many modelling situations (e.g., those which have no convenient term or process representation).  While the transformation on bigraphical reactive systems proposed by that work may give rise to a refinement that is appropriate for some situations, we aim instead in this present work to work directly within the structure of bigraphs so as to ensure generality.  As bigraphs attempt to be both a modelling formalism and a general meta-calculus for existing process calculi, it seems appropropriate that the notion of refinement we introduce should be similarly general, in the hope that we may recover calculus-specific notions of refinement within this general setting.

\section{Conclusion}

We have presented a vertical refinement mechanism for bigraphical reactive systems that adds refinement to the toolbox of model builders working within a bigraphical setting.  The addition of a sufficient condition for safe abstraction functors, and the accompanying observation that it is the \emph{preservation} of behaviour with respect to reaction that guarantees that a refinement exhibits no undesirable behaviour, provides a firm foundation from which to explore the limits and utility of this kind of vertical refinement. 

We have pointed out a clear connection to the existing work on generalising refinement across many modelling formalisms, and therefore it seems appropriate (given the application of BRSs as a \emph{meta-calculus}) that our notion of vertical refinement is also in some sense general.  We leave for future work the exploration of further mechanisms for horizontal refinement within a bigraphical setting, noting that such a notion would very likely fall within the model of general refinement, and thus likely generalise well to other modelling formalisms encoded within bigraphical reactive systems.

\bibliographystyle{eptcs}
\bibliography{refinement}

\end{document}